\documentclass[aps,twocolumn,groupedaddress]{revtex4-1}
\usepackage{graphicx}
\usepackage{xcolor}
\usepackage{hyperref}
\hypersetup{colorlinks = true, linkcolor = [rgb]{0 .4 .8}, citecolor=[rgb]{1 .2 .2}, urlcolor=black }
\usepackage{esint}
\usepackage{dsfont}
\usepackage{mathrsfs}
\usepackage{lipsum}
\usepackage{amsfonts,amssymb}
\usepackage{textpos}

\renewcommand{\thesection}{\arabic{section}}
\renewcommand{\thesubsection}{\thesection.\arabic{subsection}}
\renewcommand{\thesubsubsection}{\thesubsection.\arabic{subsubsection}}

\makeatletter
\renewcommand{\p@subsection}{}
\renewcommand{\p@subsubsection}{}
\makeatother

\makeatletter
\def\@seccntformat#1{\@ifundefined{#1@cntformat}%
{\csname the#1\endcsname\quad}
{\csname #1@cntformat\endcsname}
}
\def\@cntformat{{\normalfont\large\thesection.}\quad}
\def\subsection@cntformat{\textsection\, \thesubsection.\quad}
\def\subsubsection@cntformat{\textsection\textsection\, \thesubsubsection.\quad}
\makeatother

\newcommand{\ssection}[1]{%
\section[#1]{\normalfont\scshape #1}}
\newcommand{\ssubsection}[1]{%
   \subsection[#1]{\normalfont  #1}}

\makeatletter
\let\save@mathaccent\mathaccent
\newcommand*\if@single[3]{%
  \setbox0\hbox{${\mathaccent"0362{#1}}^H$}%
  \setbox2\hbox{${\mathaccent"0362{\kern0pt#1}}^H$}%
  \ifdim\ht0=\ht2 #3\else #2\fi
  }
\newcommand*\rel@kern[1]{\kern#1\dimexpr\macc@kerna}
\newcommand*\widebar[1]{\@ifnextchar^{{\wide@bar{#1}{0}}}{\wide@bar{#1}{1}}}
\newcommand*\wide@bar[2]{\if@single{#1}{\wide@bar@{#1}{#2}{1}}{\wide@bar@{#1}{#2}{2}}}
\newcommand*\wide@bar@[3]{%
  \begingroup
  \def\mathaccent##1##2{%
    \let\mathaccent\save@mathaccent
    \if#32 \let\macc@nucleus\first@char \fi
    \setbox\z@\hbox{$\macc@style{\macc@nucleus}_{}$}%
    \setbox\tw@\hbox{$\macc@style{\macc@nucleus}{}_{}$}%
    \dimen@\wd\tw@
    \advance\dimen@-\wd\z@
    \divide\dimen@ 3
    \@tempdima\wd\tw@
    \advance\@tempdima-\scriptspace
    \divide\@tempdima 10
    \advance\dimen@-\@tempdima
    \ifdim\dimen@>\z@ \dimen@0pt\fi
    \rel@kern{0.6}\kern-\dimen@
    \if#31
      \overline{\rel@kern{-0.6}\kern\dimen@\macc@nucleus\rel@kern{0.4}\kern\dimen@}%
      \advance\dimen@0.4\dimexpr\macc@kerna
      \let\final@kern#2%
      \ifdim\dimen@<\z@ \let\final@kern1\fi
      \if\final@kern1 \kern-\dimen@\fi
    \else
      \overline{\rel@kern{-0.6}\kern\dimen@#1}%
    \fi
  }%
  \macc@depth\@ne
  \let\math@bgroup\@empty \let\math@egroup\macc@set@skewchar
  \mathsurround\z@ \frozen@everymath{\mathgroup\macc@group\relax}%
  \macc@set@skewchar\relax
  \let\mathaccentV\macc@nested@a
  \if#31
    \macc@nested@a\relax111{#1}%
  \else
    \def\gobble@till@marker##1\endmarker{}%
    \futurelet\first@char\gobble@till@marker#1\endmarker
    \ifcat\noexpand\first@char A\else
      \def\first@char{}%
    \fi
    \macc@nested@a\relax111{\first@char}%
  \fi
  \endgroup
}
\makeatother

\usepackage{titlesec}
\titleformat{\section}{\large\bfseries\center}{\thesection.}{1em}{}
\titleformat{\subsection}{\normalsize \bfseries}{\S\, \thesubsection.}{1em}{}

\begin{document}

\newcommand{\bea}{\begin{equation}}
\newcommand{\eea}{\end{equation}}
\newcommand{\bear}{\begin{eqnarray}}
\newcommand{\eear}{\end{eqnarray}}
\newcommand{\bearr}{\begin{eqnarray*}}
\newcommand{\eearr}{\end{eqnarray*}}
\newcommand{\beal}{\begin{align}}
\newcommand{\eeal}{\end{align}}
\newcommand{\beall}{\begin{align*}}
\newcommand{\eeall}{\end{align*}}
\newcommand{\cf}{\mathcal{F}}
\newcommand{\tr}{\mathrm{tr}\,}
\newcommand{\CP}{\mathds{C}\mathds{P}}
\newcommand{\CC}{\mathds{C}}
\newcommand{\RR}{\mathds{R}}
\newcommand{\dd}{\partial}
\newcommand{\flag}{\mathcal{F}_3}

\title{\Large Integrable properties of $\sigma$-models with non-symmetric target spaces}
\author{Dmitri Bykov}
\email[]{Dmitri.Bykov@aei.mpg.de, dbykov@mi.ras.ru}
\affiliation{Max-Planck-Institut f\"ur Gravitationsphysik, Albert-Einstein-Institut, Am M\"uhlenberg 1, D-14476 Potsdam-Golm, Germany\\
Steklov
Mathematical Institute of Russ. Acad. Sci., Gubkina str. 8, 119991 Moscow, Russia}

\begin{abstract}
It is well-known that $\sigma$-models with symmetric target spaces are classically integrable. At the example of the model with target space the flag manifold $\frac{U(3)}{U(1)^3}$ -- a non-symmetric space -- we show that the introduction of torsion allows to cast the equations of motion in the form of a zero-curvature condition for a one-parametric family of connections, which can be a sign of integrability of the theory. We also elaborate on geometric aspects of the proposed model.
\end{abstract}

\maketitle

\begin{textblock}{4}(11,-4.4)
\rule{0.05pt}{1.2cm}
\hspace{-0.1cm}\rule{2.5cm}{0.05cm}\hspace{-0.1cm}
\rule{0.05pt}{1.2cm}
\end{textblock}
\begin{textblock}{4}(11.1,-4)
\emph{AEI-2014-063}
\end{textblock}

\vspace{-0.9cm}
\ssection{The setup}

A $\sigma$-model is a two-dimensional field theory describing maps $X: \Sigma \to \mathcal{M}$ from a worldsheet $\Sigma$ to a target space $\mathcal{M}$. Here we will assume that $\Sigma=\RR^2$, endowed with Euclidean metric, and that the target space is a homogeneous space, $\mathcal{M}={G\over H}$, equipped with a metric $g$. The most crucial ingredient, however, will be the torsion tensor $T^{\alpha}_{\;\beta\gamma}$ \cite{Braaten}, which is restricted by the following crucial condition: the tensor $T_{\alpha\beta\gamma}:=g_{\alpha \mu} \,T^{\mu}_{\;\beta\gamma}$ with all lower indices is totally antisymmetric (then $T$ is called \emph{skew-torsion}) and represents a closed 3-form: $dT=0$. We will as well require that the cohomology class of $T$ is trivial: $[T]=0\in H^3(\mathcal{M}, \RR)$. For this reason there exists a 2-form $\lambda$, such that $T=d\lambda$. Clearly, $\lambda$ is defined up to an addition of a closed 2-form. We therefore assume a particular choice of $\lambda$. The action of the $\sigma$-model is then given by
\bea\label{action}
\mathcal{S}=\int_\Sigma\,d^2 x\,\|\dd X\|^2_g+\int_\Sigma\,X^\ast \lambda
\eea
We will as well assume that the fields $X$ obey suitable decay conditions at infinity, so that the addition of an exact two-form to $\lambda$ does not alter the value of the action. In this case the $\lambda$'s form an affine space, whose associated vector space is $H^2(\mathcal{M},\RR)$. Note that $\lambda$ is a Kalb-Ramond field and it is not purely topological, if the torsion is nonzero. Therefore it makes a contribution to the equations of motion and, as we will see, in some special cases the resulting equations exhibit integrable properties.

\ssection{The flag manifold}

Manifolds possessing the properties described above do exist, and in this paper we will elaborate on the example of $\mathcal{M}={U(3)\over U(1)^3}:=\mathcal{F}_3$ -- the flag manifold. It can be viewed geometrically as the space of all ordered triples of orthonormal vectors in $\CC^3$, $u_1, u_2, u_3$, $u_i \circ \bar{u}_j=\delta_{ij}$, each defined up to multiplication by phase. Henceforth  the completeness relation  is used ubiquitously ($I_3$ is the $3\times 3$ identity matrix):
\bea\label{complete}
\bar{u}_1 \otimes u_1+\bar{u}_2 \otimes u_2+\bar{u}_3 \otimes u_3=I_3\;.
\eea

\ssubsection{Topological terms}

As discussed above, the space of possible $\lambda$'s entering the action (\ref{action}) is parametrized by $H^2(\mathcal{F}_3, \RR)$. As shown in \cite{BykovHaldane}, this cohomology group can be described rather directly if one notes that there exists a Lagrangian embedding
\bea\label{Lagrembed}
i: \mathcal{F}_3 \hookrightarrow \CP^2\times \CP^2 \times \CP^2
\eea
(with the product symplectic structure in the r.h.s.). The pull-backs of the three Fubini-Study forms of the $\CP^2$'s, i.e. $i^\ast(\tilde{\Omega}^{(k)}_{FS}), \,k=1, 2, 3$, generate $H^2(\mathcal{F}_3, \RR)$. However, due to the fact that the embedding $i$ is Lagrangian, there is a relation $\sum\limits_{i=1}^3 i^\ast(\tilde{\Omega}^{(k)}_{FS})=0$, so that $H^2(\mathcal{F}_3, \RR)=\RR^2$. Therefore in general $\lambda$ depends on two parameters, which characterize the truly topological terms in the action.

\ssubsection{Invariant metrics and forms}

The $U(3)$-invariant tensors on $\mathcal{F}_3$, including the invariant metrics, can be constructed from the following invariant 1-forms:
\bea\label{forms}
J_{ij}:=u_i \circ d\bar{u}_j\, .
\eea
These satisfy the relations $J_{ji}=-\bar{J}_{ij}$. The action of the stabilizer $U(1)^3$ on the $u_i$'s is as follows: $u_k\to e^{i\alpha_k(x)}\, u_k$. Clearly, the nondiagonal forms $J_{ij}, i\neq j$, transform in `bifundamental' representations, $J_{ij} \to e^{i(\alpha_i-\alpha_j)}\,J_{ij}$, whereas the diagonal ones transform as connections, $J_{kk} \to J_{kk} -i \,d\alpha_k$.

The important difference between symmetric spaces and non-symmetric ones is that the latter may possess a whole family of invariant metrics. Indeed, in the case at hand the general invariant metric can be written as
\bea
ds^2=\sum\limits_{i\neq j}\;C_{ij}\,|J_{ij}|^2,
\eea
with $C^T=C$ and $C_{ij}>0$. Thus, we see that there are three parameters in the metric, $C_{12}, C_{13}, C_{23}$. Additional requirements on the metric may reduce arbitrariness in the choice if these parameters. For example, if one insists that the metric be Einstein, there are four possible choices \cite{FlagEinsteinMetrics} (up to scaling):
\begin{itemize}
\item $(C_{12}, C_{13}, C_{23})=(1, 1, 2)$, and permutations thereof. The resulting metrics are K\"ahler-Einstein.
\item $(C_{12}, C_{13}, C_{23})=(1, 1, 1)$. In this case the metric is Einstein but not K\"ahler.
\end{itemize}
All of these metrics will be relevant for the foregoing discussion, but at the moment we wish to make a few remarks regarding the latter one, since the $\sigma$-models introduced below will be based on it. First of all, this metric is inherited from the bi-invariant metric on $SU(3)$, so that the flag manifold equipped with this metric is a \emph{naturally reductive} homogeneous space (see \cite{Agricola} for definition). Moreover, in this case it is a \emph{nearly K\"ahler} manifold (see \S~\ref{nearkahsec} below). Interestingly, it is precisely this metric that arose via the so-called Haldane limit of a spin chain in \cite{BykovHaldane}.

For the case of the flag manifold the most general `pre-torsion' two-form $\lambda$ ($d\lambda=T$), which incorporates information about the torsion and topological terms of the $\sigma$-model, can be built explicitly, using the forms (\ref{forms}). We construct the gauge-invariant (i.e. $U(1)^3$-invariant) real two-forms
\bea\nonumber
\omega_{kj}:=i J_{kj}\wedge J_{jk}=-i J_{kj}\wedge \bar{J}_{kj}
\eea
Note that $\omega_{jk}=-\omega_{kj}$, so that there are only three of them. $\lambda$ is built as a linear combination of these:
\bea\nonumber
\lambda=\frac{1}{2\pi}\,\sum\limits_{i<j}\;B_{ij}\,\omega_{ij},
\eea
with arbitrary coefficients $B_{ij}$, $B^{T}=-B$. We will see, however, that the requirement of integrability imposes a restriction on $B_{ij}$.

\ssection{The models}\label{integrmodsec}

The discussion above may be summarized by the following action:
\bear\nonumber
\mathcal{S}= \frac{1}{4\pi} \int d^2 x\, \sum\limits_{k\neq j} \,\hspace{-0.3cm}&\!\!\!\!\Big(- (J_{kj})_\mu\,(J_{jk})_\mu +  \\ \label{Lagr2} &\hspace{0.4cm}+ \, i\,B_{kj}\,\epsilon_{\mu\nu}(J_{kj})_\mu\,(J_{jk})_\nu \Big)
\eear
Here and whenever appropriate we identify the forms $J_{jk}$ with their pull-backs to the worldsheet, $X^\ast(J_{jk})$.
By construction, the action is $U(3)$-invariant, so there exists a Noether current $K$ associated to this symmetry:
\bea\label{Noether}
K=\sum\limits_{k\neq j}\;\left( J_{jk}\,-i \,B_{kj}\,\ast J_{jk}\,\right) \,\bar{u}_j \otimes u_k
\eea
Note that $K^\dagger=-K$, and the star is defined by $(\ast J)_\alpha:=\epsilon_{\alpha\beta}\,J_{\beta}$. Current conservation, $d\ast K=0$, is equivalent to the e.o.m. of the model. In order for the model to be classically integrable, the current $K$, if viewed as a connection, should be flat \cite{Eichenherr}: $dK-K\wedge K=0$. When this holds, one can construct a one-parametric family of flat connections $A$ (which ensures the compatibility of an associated Lax pair):
\bea\nonumber
A=\frac{1+\alpha}{2} \, K+\frac{\beta}{2}\,\ast K,\quad \alpha^2+\beta^2=1
\eea

Our main statement is as follows: the connection/current $K$ is flat when $B_{kj}$ satisfy the following equations:
\bear\label{Bcond}
&B_{kj}^2=1\;\;(k\neq j)\;,&\\ \nonumber & 1+B_{12}B_{23}-B_{13}B_{12}-B_{13}B_{23}=0\;.&
\eear
The solutions are easily counted:
\begin{center}
  \begin{tabular}{ | c | c | c | c | c| c| c| }
    \hline
    & (123) & (132) & (312) & (321) & (231) & (213)\\ \hline
    $B_{12}$ & 1 & 1 & 1 & -1 & -1 & -1\\ \hline
    $B_{13}$ & 1 & 1 & -1  & -1 & -1 & 1 \\ \hline
    $B_{23}$ & 1 & -1 & -1 & -1 & 1 & 1 \\ \hline
  \end{tabular}
\end{center}
The top line of the table indicates that the solutions are in one-to-one correspondence with the elements of the permutation group $\mathbf{S}_3$. Any `integrable' $\lambda$-term may be obtained from any other by permuting the three lines of the flag.
\begin{center}
$\ast\;\ast\;\ast$
\end{center}
A minor reservation is in order. The coefficients $B_{ij}$ have to be as in the table above for the current $K$ defined by (\ref{Noether}) to be flat. However, one is still free to add topological terms to the Lagrangian, i.e. closed combinations of the forms $\omega_{ij}$. Such terms formally make contributions $\delta K$ to the Noether current of the form $\delta K=\ast d M$, where $M$ is a \emph{local} function of the fields. Therefore $\delta K$ is conserved regardless of the e.o.m. As such, it may safely be omitted from the Noether current, since otherwise it would ruin its flatness. To summarize, additional topological terms $\omega_{12}+\omega_{13}$, $\omega_{13}+\omega_{23}$ (and linear combinations thereof) may be introduced, but their contributions should not be taken into account in $K$. 
\begin{center}
$\ast\;\ast\;\ast$
\end{center}

To prove that $K$ is flat we rewrite the current (\ref{Noether}) in simpler form, forming a matrix $g$ out of the three vectors $u_i$, i.e. $(u_i)_s\equiv g_{is}, (\bar{u}_j)_p\equiv (g^\dagger)_{pj}$. Then
\bea\nonumber
K=g^\dagger\circ \underbrace{\left( \begin{array}{ccc}
0 & P_{12}^+\,J_{12} & P_{13}^+ \,J_{13} \\
P_{12}^-\,J_{21} & 0 & P_{23}^+ \,J_{23} \\
P_{13}^- \, J_{31} & P_{23}^-\,J_{32} & 0 \end{array} \right)}_{:=S} \circ \;g\;,
\eea
where $P_{mn}^{\pm}:=1\pm i \,B_{mn}\,\ast$ satisfy $(P_{mn}^{\pm})^2=2\,P_{mn}^{\pm}$.

Using (\ref{complete}), one can verify that the following holds true: $dg=-J\circ g,\; dg^\dagger=g^\dagger \circ J$ ($J$ is a matrix with components $J_{ij}$). The current conservation condition $d\ast K=0$ and the flatness condition $dK-K\wedge K=0$ may then be reformulated respectively as
\bear\nonumber
&d\ast S+\{J, \ast S\}=0&\\ \nonumber
&dS+\{J, S\}-S\wedge S=0&
\eear
One can check directly that these two equations are \emph{equivalent} for the matrix $S$ above, if the conditions (\ref{Bcond}) are fulfilled. In the course of the calculation the following relations are useful: $\ast^2=-1, (\ast a) \wedge b=-a\wedge (\ast b)$.

For the moment let us focus on the case $B_{12}=B_{13}=B_{23}=1$. Upon the introduction of the complex coordinate $z:=x^1+i \,x^2$ the Lagrangian of (\ref{Lagr2}) can be written in a much more compact form:
\bear\label{Lagr3}\nonumber
\mathcal{L}=&&\!\!\!\!\!(u_1\circ \dd_{\bar{z}} \bar{u}_2)\,(\bar{u}_1\circ \dd_{z} u_2)+(u_1\circ \dd_{\bar{z}} \bar{u}_3)\,(\bar{u}_1\circ \dd_{z} u_3)+\\  &+&(u_2\circ \dd_{\bar{z}} \bar{u}_3)\,(\bar{u}_2\circ \dd_{z} u_3)
\eear
This is a direct generalization of the Lagrangian for the target space $\CP^1=\frac{U(2)}{U(1)^2}$, which can be written as
\bea\label{LagrSphere}
\mathcal{L}_{\CP^1}=(u_1\circ \dd_{\bar{z}} \bar{u}_2)\,(\bar{u}_1\circ \dd_{z} u_2)\; .
\eea
An important fact is that in the Lagrangian (\ref{Lagr3}) the complex structure on the worldsheet is correlated with the complex structure in the space $(u(1)^3)^{\perp}\subset u(3)$ -- the space of anti-Hermitian off-diagonal matrices. For example, in (\ref{Lagr3}) there is a term $(u_1\circ \dd_{\bar{z}} \bar{u}_2)\,(\bar{u}_1\circ \dd_{z} u_2)$ but no counterpart with $z \leftrightarrow \bar{z}$. In \S~\ref{compstructsec} we will see that the models defined by the action (\ref{Lagr2}) with different values of $B_{kj}$ are in fact related to the choice of complex structure on the flag manifold~$\flag$.

\ssubsection{Local conserved charges}

It is well-known that integrability requires, a la Liouville, the existence of an infinite number of commuting conserved charges. Therefore a  graphic way of checking the integrability of the model is to directly build an infinite sequence of conserved charges using the equations of motion, i.e. the current conservation equation $d\ast S+\{J, \ast S\}=0$. It reduces to the following equations for the components $J_{ij}$:
\bear\label{eom}
&D[(1-i\ast) J_{21}]=0,\quad D[(1-i\ast) J_{32}]=0,&\\ \label{eom2} &D[(1-i\ast ) J_{31}]+2\, J_{32}\wedge J_{21}=0&
\eear
and their complex conjugates. Here $D$ is the covariant derivative for the group $U(1)^3$, i.e.
\bea \nonumber
D J_{kj}:=dJ_{kj}+(J_{kk}-J_{jj})\wedge J_{kj}.
\eea
It turns out that the equation (\ref{eom2}) can be transformed in a rather remarkable way, if one uses the relation (\ref{complete}):
\bea \nonumber
J_{32}\wedge J_{21}=-du_3\circ \hspace{-0.6cm}\underbrace{\;\bar{u}_2 \wedge u_2 \;}_{=I_3-\bar{u}_1 \otimes u_1-\bar{u}_3\otimes u_3}\hspace{-0.6cm}\circ \;d\bar{u}_1=-DJ_{31}
\eea
Therefore the e.o.m. above can be rewritten in a much more symmetric form:
\bear \nonumber
D[(1-i\ast) J_{21}]=0,\;\;\;&&\;\;\; D[(1-i\ast) J_{32}]=0, \\ \nonumber 
D[(1+i\ast) J_{31}]=0,\;\;\;&&\;\;\; D[(1+i\ast) J_{12}]=0,\\  \nonumber
D[(1+i\ast) J_{23}]=0,\;\;\;&&\;\;\; D[(1-i\ast) J_{13}]=0.
\eear
Noting that $(1-i\ast) J=(J_1-i J_2)\,d(x^1+i x^2):=J_{z}\,dz$, we can derive from the above equations the holomorphic conservation law:
\bea \nonumber
\dd_{\bar{z}}\left((J_{13})_z (J_{32})_z (J_{21})_z\right)=0
\eea
The gauge-invariant quantity $H:=(J_{13})_z (J_{32})_z (J_{21})_z$ generates an infinite number of conservation laws, since $\dd_{\bar{z}}(H^n)=0$ for $n=1, 2, \ldots$. Note that the $\sigma$-model described by the Lagrangian (\ref{Lagr2}) has the energy momentum-tensor
\bea \nonumber
T_{zz}=(J_{12})_z\,(J_{21})_z+(J_{13})_z\,(J_{31})_z+(J_{23})_z\,(J_{32})_z,
\eea
which is holomorphic as well: $\dd_{\bar{z}}T_{zz}=0$ (To check this one needs to rewrite some of the e.o.m. in the `elongated' form (\ref{eom2})). However, $H$ is a holomorphic current that is independent of the energy-momentum tensor and hence not directly related to the classical conformal invariance of the theory.

\ssection{Geometry of the flag manifold}

In this section we will try to understand certain aspects of the action (\ref{Lagr2}) and, in particular, the variety of the allowed values of $B_{mn}$ summarized in the table above. To this end, certain facts about complex structures on the flag manifold will be of importance to us.

\ssubsection{Complex structures}\label{compstructsec}

The most fundamental fact is that there are 8 invariant almost complex structures on $\mathcal{F}_3$. Rather concretely, they may be defined as follows: pick the three basic 1-forms $J_{12}, J_{13}, J_{23}$ and \emph{postulate} that each of these is either holomorphic or antiholomorphic. Then pick a basis of the chosen holomorphic 1-forms $J_{k}$, where $k$ can stand for $12, 13, 23$ or their conjugates. In order for the almost complex structure to be integrable, the holomorphic 1-forms should constitute a differential ideal, i.e. $dJ_{k}=\sum\limits_{m} \,a_{km} \wedge J_m$ for some coefficient 1-forms $a_{km}$. Using the identity $dJ_{ij}=-\sum\,J_{ik}\wedge J_{kj}$, one finds that differential ideals are formed by the following triples of 1-forms (plus the conjugate ones):
\begin{itemize}
\item $I_1=\{ J_{12}, \;J_{13},\;J_{23}\}$
\item $I_2=\{ J_{12}, \;J_{31},\; J_{32}\}$
\item $I_3=\{ J_{21},\; J_{31},\;J_{23}\}$
\end{itemize}
On the contrary, the almost complex structure defined by $\{ J_{12}, \;J_{31},\;J_{23}\}$ (and the conjugate one) is not integrable. Looking at the action (\ref{Lagr3}), which corresponds to the choice $B_{12}=B_{13}=B_{23}=1$, one realizes that the absolute minima of the action correspond to $I_1$-\emph{holomorphic} curves:
\bea \nonumber
u_1\circ \dd_{\bar{z}} \bar{u}_2=0,\quad u_1\circ \dd_{\bar{z}} \bar{u}_3=0,\quad u_2\circ \dd_{\bar{z}} \bar{u}_3=0
\eea
Accordingly, the choice $B_{12}=1, B_{13}=B_{23}=-1$ leads to $I_2$-holomorphic curves, and $B_{12}=B_{13}=-1$, $B_{23}=1$ leads to $I_3$-holomorphic curves as minima of the action. We now claim that the three actions differ by topological terms, which in any given topological sector are field-independent constants. This has important consequences for the existence of holomorphic curves. Furthermore, this implies that the e.o.m. of the three models are the same.

\ssubsection{$I_1, I_2, I_3$-holomorphic curves}

In order to prove the claim that we have made concerning the actions (\ref{Lagr2}) with different values of $B$'s we note that the exterior derivative of the Kalb-Ramond form $\lambda= \frac{i}{2\pi}\,(B_{12}\,J_{12}\wedge J_{21}+B_{13}\,J_{13}\wedge J_{31}+B_{23}\,J_{23}\wedge J_{32})$ is
\bea \nonumber
d\lambda=\frac{i}{2\pi}\,(B_{23}+B_{12}-B_{13})\,(J_{21}\wedge J_{13}\wedge J_{32} - \mathrm{c.c.})
\eea
In particular, it is the same for the $B$'s in the 1, 3, 5 and 2, 4, 6 columns of the table above. This means that the corresponding Kalb-Ramond forms differ by closed, i.e. topological, 2-forms. Let us check this explicitly. Denoting the Kalb-Ramond form corresponding to the values of $B$ in the $k$-th column of the table as $\lambda^{(k)}$, we find:
\bear \nonumber
\lambda^{(1)}-\lambda^{(3)}&=&2\,\Omega_{FS}^{(3)}\\ \nonumber
\lambda^{(3)}-\lambda^{(5)}&=&2\,\Omega_{FS}^{(2)}\\ \nonumber
\lambda^{(5)}-\lambda^{(1)}&=&2\,\Omega_{FS}^{(1)}
\eear
Here $\Omega_{FS}^{(p)}=i^\ast(\tilde{\Omega}^{(p)}_{FS})$, where $i$ is the embedding (\ref{Lagrembed}), the Fubini-Study form $\tilde{\Omega}_{FS}$ is
$$
\tilde{\Omega}_{FS}~=~\frac{i}{2\pi}\left( du\wedge\circ\,d\bar{u}-(du\circ\bar{u})\wedge (d\bar{u}\circ u)\right)$$ and $\sum\limits_{p=1}^3\,\Omega_{FS}^{(p)}=0$. We will denote by $n_p\in \mathds{Z}$ the integral of the pull-back of the corresponding Fubini-Study form over the worldsheet, i.e. $n_p=\int\limits_{\Sigma}\,\Omega_{FS}^{(p)}$. These are subject to the condition $\sum\limits_{p=1}^3\,n_p=0$. Therefore for the difference in the actions, corresponding to the Kalb-Ramond forms~$\lambda^{(k)}$, which we analogously call $\mathcal{S}^{(k)}$, we obtain:
\bear \nonumber
\mathcal{S}^{(1)}-\mathcal{S}^{(3)}&=&2\,n_3\\ \nonumber
\mathcal{S}^{(3)}-\mathcal{S}^{(5)}&=&2\,n_2\\ \nonumber
\mathcal{S}^{(5)}-\mathcal{S}^{(1)}&=&2\,n_1
\eear
Now, suppose there exists an $I_1$-holomorphic curve. In this case $\mathcal{S}^{(1)}=0$. On the other hand, all of the actions are nonnegative: $S^{(k)}\geq 0$, so we obtain the necessary condition:
\bea\label{bound1}
I_1:\quad\quad n_1\geq 0,\quad n_3\leq 0
\eea
Analogously the $I_2$ and $I_3$-holomorphic curves require
\bear\label{bound2}
I_2:\quad\quad n_2\leq 0,\quad n_3\geq 0\\ \label{bound3}
I_3: \quad\quad n_1\leq 0,\quad n_2\geq 0
\eear

\begin{figure}[h]
    \centering
    \includegraphics[width=0.6 \columnwidth]{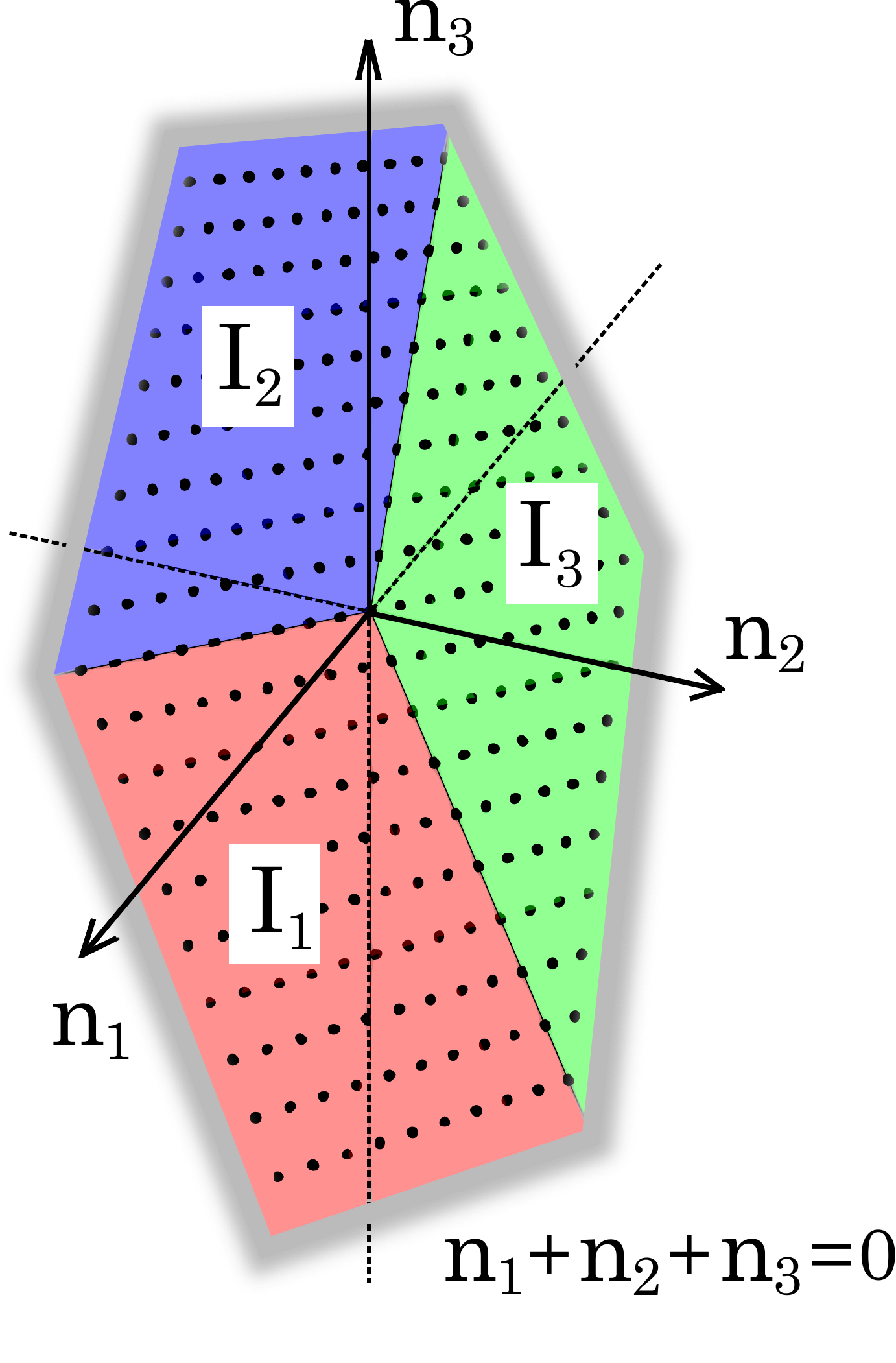}
    \caption{Admissible instanton numbers for various complex structures on $\flag$ are shown by black dots.}
    \label{instnumfig}
\end{figure}

It follows that generically a curve with given topological numbers can only be holomorphic in one of the complex structures (see Fig. \ref{instnumfig}). The only exception is when one of the numbers $n_p$ vanishes. Assume for the moment that $n_3=0$ and we are dealing with a $I_1$-holomorphic curve. Writing out explicitly the definition of $n_3$,
\bear \nonumber
n_3&=&\frac{i}{2\pi}\,\int\,\Big(J_{13}\wedge J_{31}+J_{23}\wedge J_{32}\Big)=\\ \nonumber
&=&\frac{i}{2\pi}\,\int\,\Big((J_{13})_z (J_{31})_{\bar{z}}-(J_{31})_z (J_{13})_{\bar{z}} + \\ \nonumber  &&\;\;\;\;\;\;\;+\, (J_{23})_z (J_{32})_{\bar{z}}-(J_{32})_z (J_{23})_{\bar{z}}\Big)\,dz\wedge d\bar{z}\;,
\eear
and recalling that for a $I_1$-holomorphic curve $(J_{13})_{\bar{z}}=(J_{23})_{\bar{z}}=0$, we see that $n_3=0$ implies $(J_{13})_{z}=(J_{23})_{z}=0$, so that the curve is $I_2$-holomorphic as well. Such a curve can be described rather explicitly. Indeed, $(J_{13})_{\bar{z}}=(J_{23})_{\bar{z}}=0$ implies that $\dd_{\bar{z}}\bar{u}_3$ is orthogonal to both $u_1$ and $u_2$ and hence proportional to $\bar{u}_3$, i.e. $\dd_{\bar{z}}\bar{u}_3~=~\alpha\,\bar{u}_3$. Analogously $I_2$-holomorphicity implies $\dd_z \bar{u}_3=\beta \bar{u}_3$. Compatibility of these equations requires $\alpha={\dd \phi \over \dd \bar{z}}$ and $\beta={\dd \phi \over \dd z}$ for some function $\phi(z, \bar{z})$, and the solution then takes the form $\bar{u}_3(z, \bar{z})=e^{\phi(z, \bar{z})}\,\bar{u}_3^{(0)}$, where $\bar{u}_3^{(0)}$ is a constant unit vector. From the normalization of $u_3$, $u_3 \circ \bar{u}_3=1$, it follows that $\phi$ is purely imaginary, so that $\bar{u}_3(z, \bar{z})$ is a gauge transformation of $\bar{u}_3^{(0)}$. This means that the curve, which is $I_1$ and $I_2$-holomorphic, maps trivially to the third line of the flag, i.e. it is essentially a map to the $\CP^1$ parametrized by $u_1, u_2$ with fixed $u_3$. It will be explained in \S\, \ref{twistorsec} below that this $\CP^1$ is the fiber of one of the forgetful bundles (\ref{fiberbundle}).

To summarize, a curve is holomorphic with respect to two of the complex structures at the same time if and only if it is a map to a \emph{fiber} of one of the bundles (\ref{fiberbundle}). Such curves are therefore labeled by points of the base of the bundle, i.e. of the $\CP^2$ (in the case above this $\CP^2$ is parametrized by the vector $u_3$).
\vspace{0.5cm}
\begin{center}
$\ast\;\ast\;\ast$
\end{center}
This discussion has a more basic parallel for the case of instantons/anti-instantons in the K\"ahler metric on $\flag$. In general, for a K\"ahler target space, the $\sigma$-model actions for the two opposite complex structures,
\bea \nonumber
\tilde{\mathcal{S}}^{(1)}=i\,\int\,dz\wedge d\bar{z}\,\left(G_{i\bar{j}}\,\dd_z X^i\,\dd_{\bar{z}}\bar{X}^{\bar{j}}\right)
\eea
and
\bea \nonumber
\tilde{\mathcal{S}}^{(2)}=i\,\int\,dz\wedge d\bar{z}\,\left(G_{i\bar{j}}\,\dd_{\bar{z}} X^i\,\dd_{z}\bar{X}^{\bar{j}}\right),
\eea
differ by the integral of the pull-back of the K\"ahler form:
\bea \nonumber
\tilde{\mathcal{S}}^{(1)}-\tilde{\mathcal{S}}^{(2)}=\,\int\,X^\ast(G),
\eea
where $G=i\,G_{i\bar{j}}\,dX^i\wedge d\overline{X}^{\bar{j}}$ is the \emph{real} K\"ahler form. Note that $\tilde{\mathcal{S}}^{(1)}\geq 0$ and $\tilde{\mathcal{S}}^{(2)}\geq 0$. For a holomorphic curve~$\Sigma$, $\tilde{\mathcal{S}}^{(2)}=0$, so that non-negativity of $\tilde{\mathcal{S}}^{(1)}$ requires that $\int\limits_\Sigma\,X^\ast(G)\geq 0$. One can check that the K\"ahler forms $G_1, G_2, G_3$ for the complex structures $I_1, I_2, I_3$ are expressed in terms of the Fubini-Study forms $\Omega_{FS}^{(p)}$ as
\bear \nonumber
G_1=\Omega_{FS}^{(1)}-\Omega_{FS}^{(3)}\\ \nonumber
G_2= \Omega_{FS}^{(3)}-\Omega_{FS}^{(2)}\\ \nonumber
G_3=\Omega_{FS}^{(2)}-\Omega_{FS}^{(1)}
\eear
The requirement $\int\limits_\Sigma\,X^\ast(G)\geq 0$ then leads to the following conditions for the $I_1, I_2, I_3$-holomorphic curves:
\bea \nonumber
I_1: n_1 \geq n_3,\quad\quad
I_2: n_3 \geq n_2,\quad\quad
I_3: n_2 \geq n_1
\eea
These are weaker bounds than (\ref{bound1})-(\ref{bound3}), therefore they are automatically satisfied for the points in Fig. \ref{instnumfig}.

\ssubsection{The flag manifold as a twistor space}\label{twistorsec}

Now that we have seen that the complex structures on $\flag$ are of utmost importance for the $\sigma$-models introduced in section \ref{integrmodsec},  we wish to take yet another perspective at these complex structures. The most relevant fact is that the flag manifold is a twistor space of the complex projective plane $\overline{\CP^2}$ (the $\CP^2$ with reversed orientation) \cite{AHS}, \cite{Salamon}. It turns out that all of the invariant almost complex structures on $\flag$ may be constructed as natural almost complex structures of the twistor space.

We will parametrize the complex projective plane $\CP^2$ by a unit vector $u_2\in \CC^3$, defined up to multiplication by a phase. Given a point in $\CP^2$, i.e. a vector $u_2$, pick two unit vectors, $u_1$ and $u_3$, orthogonal to each other and to $u_2$. Then the cotangent space to $\CP^2$ at $u_2$ is spanned by the 1-forms $u_1 \circ d \bar{u}_2, u_3 \circ d\bar{u}_2, \bar{u}_1 \circ du_2, \bar{u}_3 \circ du_2$. In order to choose a complex structure in $T^\ast_{u_2}\CP^2$, we pick two of these one-forms, which we call $J_1$ and $J_2$, and postulate that they are holomorphic (the other two therefore being anti-holomorphic). Note, however, that the choice has to be compatible with hermiticity of the metric and with the (reversed) orientation of $\CP^2$, meaning that the value of the square of the corresponding K\"ahler form on any quadruple of vectors should be of opposite sign to the value of the square of the Fubini-Study form. One readily sees that the choice $J_1:=u_1 \circ d\bar{u}_2$ and $J_2:=du_2 \circ \bar{u}_3$ is admissible. Changing the vectors $u_1, u_3$, while preserving $u_2$, corresponds to changing the complex structure in $T^\ast_{u_2}\CP^2$.
Indeed, any two pairs $(u_1 ,u_3)$ and $(u_1', u_3')$ are connected by a basis rotation
\bea \nonumber
\left(\begin{array}{c}
u_1'\\ u_3'
\end{array}\right)=\Xi \circ \left(\begin{array}{c}
u_1\\ u_3
\end{array}\right),\quad \Xi \in SU(2),
\eea
since this is the transformation preserving the orthonormality relations $\bar{u}_1 \circ u_1=\bar{u}_3 \circ u_3=1, \bar{u}_1 \circ u_3=0$. This transformation has the effect of rotating the one-forms $(J_1, \overline{J_2})$:
\bea \nonumber
\left(\begin{array}{c}
J_1\\ \overline{J_2}
\end{array}\right) \to\Xi \circ \left(\begin{array}{c}
J_1\\ \overline{J_2}
\end{array}\right),\quad \Xi \in SU(2),
\eea
which changes the complex structure unless $\Xi\in U(1)$. Therefore the space of such complex structures is isomorphic to $\frac{SU(2)}{U(1)}=\CP^1$. This is the fiber of the twistor fiber bundle $\mathrm{Tw}(\CP^2)$. Note as well that the above transformation leaves the Fubini-Study metric on $\CP^2$ unchanged, since the latter can be written as
\bea \nonumber
(ds^2)_{\CP^2}=du_2 \circ d\bar{u}_2 -|u_2 \circ d\bar{u}_2|^2=|J_1|^2+|J_2|^2\,.
\eea

As we have seen, a point in the twistor space is given by a triplet of orthonormal vectors $u:=(u_1, u_2, u_3)$, defined up to phases, which means that $\mathrm{Tw}(\CP^2)\simeq \flag$. The cotangent space at this point is $T^\ast_{u}\mathrm{Tw}(\CP^2)=T^\ast_{u_2}\CP^2\oplus T^\ast_{(u_1, u_3)} \CP^1$. We wish to define a complex structure on this space. Since we have already defined the complex structure on $T^\ast_{u_2}\CP^2$ at a point $(u_1, u_3)$ in the fiber of the twistor fiber bundle, we need only define the complex structure in the fiber directions, i.e. on $T^\ast_{(u_1, u_3)} \CP^1$. The cotangent space to this $\CP^1$ is spanned by the 1-forms $u_1 \circ d \bar{u}_3, u_3 \circ d \bar{u}_1$. We may declare either one of them to be holomorphic, thereby introducing \emph{two} natural almost complex structures on the twistor space. (We note in passing that this procedure generalizes directly to twistor spaces of other manifolds.) For the time being let us declare $u_1 \circ d \bar{u}_3$ to be the holomorphic one. Then we obtain the complex structure on $\flag$, in which the forms $u_1 \circ d\bar{u}_2, du_2 \circ \bar{u}_3, u_1 \circ d \bar{u}_3$ are holomorphic. Clearly, this is the complex structure $I_1$ discussed before. Had we chosen the form $u_3 \circ d \bar{u}_1$ to be the holomorphic one-form cotangent to the fiber, we would have arrived at the non-integrable almost complex structure $\tilde{I}$. Similarly to the integrable ones, it may be defined by the triple of holomorphic one-forms:
\bea\label{nonintcs}
\tilde{I}=\{ J_{12}, \;J_{31},\;J_{23}\}
\eea
To obtain the complex structures $I_2$ and $I_3$ in a similar fashion, recall that there are \emph{three} forgetful projections
\vspace{-0.5cm}
\bea\label{fiberbundle}
\arraycolsep=1.4pt\def\arraystretch{2.2}
\begin{array}{ccc}
&\flag&  \\
\hspace{0.3cm}\CP^1 &\left\downarrow\rule{0cm}{0.7cm}\right.&\!\!\!\!\!\pi_1, \pi_2, \pi_3  \\
&\CP^2 &
\end{array}
\vspace{-0.3cm}
\eea
which unite two of the three lines of the flag into a plane. In other words, we can view the same flag manifold as a twistor space for the projective planes parametrized by $u_1$ or $u_3$, not just $u_2$. In this case, however, the twistor space structure imposes different complex structures on $\flag$: $I_2$ and $I_3$, respectively. On the other hand, reversing the complex structure in the fiber no longer produces any new complex structures and leads us back to the non-integrable almost complex structure $\tilde{I}$ or its opposite.

\ssubsection{The nearly K\"ahler structure}\label{nearkahsec}

Here we wish to demonstrate that $\flag$ is nearly K\"ahler for the metric
\bea\label{projmetric}
ds^2=|J_{12}|^2+|J_{13}|^2+|J_{23}|^2,
\eea
which we have used in the action (\ref{Lagr2}), and the non-integrable almost complex structure (\ref{nonintcs}). By definition, `nearly K\"ahler' means that the covariant derivative of the complex structure tensor with lowered indices (the K\"ahler form), i.e. $\nabla_\alpha\,I_{\mu\nu}$, is completely skew-symmetric (for the properties of nearly K\"ahler manifolds see~\cite{Gray}).

The key property of the metric (\ref{projmetric}) is that it is induced from the bi-invariant metric on $SU(3)$. In view of this it will be useful to recall the decomposition $su(3)=u(1)^2\oplus [u(1)^2]^\perp$. Having in mind a more general setup when $SU(3)$ is replaced by a different Lie group $G$, we will use the corresponding notation $\mathfrak{g}=\mathfrak{h}\oplus \mathfrak{m}$.

Restating the nearly K\"ahler condition in a local basis, we need to prove that $E^\alpha_a\,\nabla_\alpha\,I_{bc}$ is skew-symmetric in $a, b, c$ (where $E^\alpha_a$ is the inverse vielbein). Note that in a local basis the invariant almost complex structure $I_{bc}$ is constant. Apart from that, since the metric is induced from the bi-invariant metric on a group, the spin connection $\omega$ in a local basis is proportional to the structure constants, and $E^\alpha_a (\omega_\alpha)_{bc}\propto f_{abc}$ (here the indices are restricted to the space $\mathfrak{m}$). Therefore the goal is to show that the combination $f_{abm} I_{mc}+f_{acm} I_{bm}$ is skew-symmetric in $a, b, c$. This combination can be also rewritten as $\tr(T_a[T_c, I(T_b)]-T_a[T_b, I(T_c)])$. Requiring antisymmetry with respect to $a \leftrightarrow b$, one obtains the condition $\tr(T_c[T_a, I(T_b)])=-\tr(T_c[T_b, I(T_a)])$ for $T_a, T_b, T_c \in \mathfrak{m}$, which implies
\bea \nonumber
[T_a, I(T_b)]-[I(T_a), T_b] \in \mathfrak{h}
\eea
It can be restated as follows: for any two generators $T$ and $\tilde{T}$ belonging to different eigenspaces of the operator $I$ (i.e. $I(T)=\pm i T,\;I(\tilde{T})=\mp i \tilde{T}$) their commutator should belong to $\mathfrak{h}$: $[T, \tilde{T}] \in \mathfrak{h}$. For the case at hand one can check this explicitly, using the definition of the complex structure (\ref{nonintcs}).

\ssection{Outlook}

We have seen that the introduction of a torsion term could lead to the integrability of a $\sigma$-model with a non-symmetric homogeneous target space. The most important question is to determine the class of target spaces, for which this can happen. As we pointed out, almost complex structures on the spaces in question play an important role, therefore it is natural to conjecture that integrability is related to the \emph{nearly K\"ahler} property of the space. If this is so, one should expect to discover integrable properties in the $\sigma$-models on the twistor spaces of various symmetric spaces \cite{Salamon} (which are themselves homogeneous spaces), as well as on $S^3 \times S^3$ and $S^6$. A discussion of the relevant geometric aspects of six-dimensional homogeneous nearly K\"ahler manifolds (of which $\mathcal{F}_3$ is an example) can be found in \cite{Butruille}.

A physical drawback of the model (\ref{Lagr2}) is that apparently its quantized version is non-unitary. This is so because the Kalb-Ramond term in (\ref{Lagr2}) is real in Euclidean signature, which means that it is imaginary in Minkowski signature. Therefore the action is complex in Minkowski signature.  A similar issue has been encountered in the context of topological $\sigma$-models with non-K\"ahler target spaces~\cite{Witten}. On the other hand, classically the action (\ref{Lagr2}) is well-defined and leads to a well-posed variational problem. We have described above some of its solutions (the holomorphic curves).

Apart from these foremost questions, there are several other directions, in which the results of the present paper could be extended. If the model discussed above can be consistently quantized, a natural question is whether the quantized version inherits the integrable properties (in certain cases, such as in the $\CP^N$ model, integrability does not survive quantization). One could as well inquire if there exists a supersymmetric extension of the model. It would also be interesting to see whether introduction of torsion has a bearing on the classification of integrable string $\sigma$-models (see \cite{ZaremboSym}). Finally, it is curious to find out, whether the model (\ref{Lagr2}) and its possible generalizations to other target spaces are related to gauged WZNW models of some sort \cite{Kounnas}. We believe, however, that the model (\ref{Lagr2}) is not conformal after quantization.
\begin{center}
\line(1,0){200}
\end{center}
\vspace{-0.1cm}
\appendix
\ssection{K\"ahler structures on $\mathcal{F}_3$}

Here we provide some background information on the K\"ahler metrics on $\mathcal{F}_3$, i.e. we will elaborate on the \emph{integrable} complex structures of $\mathcal{F}_3$. Since these are interchanged by the permutations of the lines of the flag, we will pick one particular complex structure, corresponding to the choice $B_{12}=B_{13}=B_{23}=1$ above, or equivalently to the choice of $\{J_{12}, J_{13}, J_{23}\}$ as a triplet of holomorphic 1-forms.

Above we constructed the most general invariant metric on $\flag$:
\bea\label{invmetr}
ds^2=\sum\limits_{i\neq j}\;C_{ij}\,|J_{ij}|^2,
\eea
with positive constants $C_{ij}$. In the chosen complex structure the metric (\ref{invmetr}) is Hermitian with the associated K\"ahler form
\bea\label{Kahform}
\Omega=C_{12}\,J_{12}\wedge J_{21}+C_{13}\,J_{13}\wedge J_{31}+C_{23}\,J_{23}\wedge J_{32}
\eea
The metric is K\"ahler when $\Omega$ is closed, leading to the condition
\bea \nonumber
C_{12}-C_{13}+C_{23}=0,
\eea
which can be solved as
\bea \nonumber
C_{12}=\lambda_1-\lambda_2,\quad C_{13}=\lambda_1-\lambda_3,\quad C_{23}=\lambda_2-\lambda_3
\eea
for some constants $\lambda_1, \lambda_2, \lambda_3$. Introducing a diagonal matrix $\lambda=\mathrm{Diag}(\lambda_1, \lambda_2, \lambda_3)$, the K\"ahler form (\ref{Kahform}) takes the shape of the Kirillov symplectic form on the adjoint orbit of $SU(3)$:
\bea \nonumber
\Omega=\tr\left(\lambda\,j\wedge j\right)\quad\textrm{with}\quad j=-g^{-1}dg \,.
\eea
We see that the space of K\"ahler metrics on $\mathcal{F}_3$, up to scaling, is one-dimensional. The Einstein condition further completely fixes the remaining parameter, so that, up to scaling, $C_{12}=C_{23}=1, C_{13}=2$.

To build K\"ahler potentials for the above metrics, we will take a holomorphic, rather than unitary, viewpoint and consider $\flag$ as the quotient $\flag=GL(3, \CC)/B$, where $B$ is the Borel subgroup of upper triangular matrices. We introduce the nondegenerate matrix
\bea\label{complexflag}
U=\left( \begin{array}{ccc}
v_1 & w_1 & z_1 \\
v_2 & w_2 & z_2 \\
v_3 & w_3 & z_3 \end{array} \right)\;,
\eea
where $v, w, z$ are three linearly independent vectors in $\CC^3$ parametrizing the flag.

Denoting by $M_{abc\ldots|mnp\ldots}$ the minor corresponding to the lines $abc\ldots$ and columns $mnp\ldots$ of the matrix $U$, the K\"ahler potential can be written as
\bear \nonumber
\mathcal{K}&=&C_{12}\,\log{\left(\sum\limits_a|M_{a|1}|^2\right)}+C_{23}\,\log{\left(\sum\limits_{a\neq b} |M_{ab|12}|^2\right)}+\\ \label{Kahflag} &+&D\,\log{\left(\sum\limits_{a\neq b \neq c} |M_{abc|123}|^2\right)}
\eear
with arbitrary coefficients $C_{12}, C_{23}, D$. Under the action of the Borel group all of the minors under each logarithm are multiplied by the same function, thereby leading to a change in the K\"ahler potential that is a sum of holomorphic and antiholomorphic functions. The construction is generalized in an obvious way to flag manifolds in any dimension. Furthermore, for the case at hand, the last term in (\ref{Kahflag}) is in fact proportional to $\log(M_{123|123})+ \log(\overline{M_{123|123}})$, so it does not contribute to the metric and can be neglected as well.

We arrive at the following K\"ahler potential:
\bea\label{Kahpotflag}
\mathcal{K}=C_{12} \,\log\|v\|^2+C_{23}\, \log\left(\|v\|^2\,\|w\|^2-|v\circ \bar{w}|^2\right)
\eea
with arbitrary constants $C_{12}>0, C_{23}>0$. This potential is gauge-invariant with respect to the action of the Borel group on (\ref{complexflag}) and, therefore, one may pick a particular gauge to remove the redundancy. One option is to pick a holomorphic gauge (similar to passing to inhomogeneous coordinates in a projective space), however to make contact with the metric written in the form (\ref{invmetr}) one should pass to unitary gauge. This amounts to assuming that the vectors $v:=u_1, w:=u_2, z:=u_3$ of (\ref{complexflag}) are orthonormal: $u_i \circ \bar{u}_j=\delta_{ij}$. In this case the metric arising from the K\"ahler potential (\ref{Kahpotflag}) can be written as
\bea \nonumber
ds^2=C_{12}\,|j_{12}|^2+C_{23}\,|j_{23}|^2+(C_{12}+C_{23})\,|j_{13}|^2
\eea
As discussed in the paper, the metric is Einstein only when $C_{12}=C_{23}$.

\vspace{-0.2cm}
\begin{acknowledgments}
I would like to thank S.~Frolov and K.~Zarembo for comments on the manuscript. I am indebted to Prof.~A.A.Slavnov and to my parents for support and encouragement. My work was supported in part by grants RFBR 14-01-00695-a, 13-01-12405 ofi-m2 and the grant MK-2510.2014.1 of the President of Russia Grant Council.
\end{acknowledgments}


\begin{thebibliography}{100}
\bibitem{Braaten}
  E.~Braaten, T.~L.~Curtright and C.~K.~Zachos,
  \emph{``Torsion and Geometrostasis in Nonlinear Sigma Models''},
  Nucl.\ Phys.\ B {\bf 260} (1985) 630.
\bibitem{BykovHaldane} 
  D.~Bykov,
 \emph{ ``Haldane limits via Lagrangian embeddings''},
  Nucl.\ Phys.\ B {\bf 855}, 1 (2012) 100,
  arXiv:1104.1419
\bibitem{FlagEinsteinMetrics}
  A. Arvanitoyeorgos,
\emph{``New invariant Einstein metrics on generalized flag manifolds''},
 Trans.\ Am.\ Math.\ Soc. {\bf 337}, 2 (1993) 981
\bibitem{Agricola}
I. Agricola, A. C. Ferreira, T. Friedrich,
\emph{``The classification of naturally reductive homogeneous spaces in dimensions $n \leq 6 $''},
(2014), arXiv:1407.4936
\bibitem{Eichenherr}
  H.~Eichenherr and M.~Forger,
  \emph{``On the Dual Symmetry of the Nonlinear Sigma Models''},
  Nucl.\ Phys.\ B {\bf 155} (1979) 381.
\bibitem{AHS}
M. F. Atiyah, N. J. Hitchin, I. M. Singer,
   \emph{``Self-duality in four-dimensional Riemannian geometry''},
    Proc.\ R. \ Soc.\ Lond., Ser. A {\bf 362} (1978) 425
\bibitem{Salamon}
S. Salamon,
    \emph{``Harmonic and holomorphic maps''},
Geometry Semin. ''Luigi Bianchi'', Lect. Sc. Norm. Super., Pisa (1984), Lect. Notes Math. 1164 (1985) 161
\bibitem{Gray}
A. Gray,
    \emph{``Nearly K\"ahler manifolds''},
J. Differ. Geom. {\bf 4} (1970) 283
\bibitem{Butruille}
J.-B. Butruille,
\emph{``Classification des vari\'{e}t\'{e} approximativement k\"{a}hleriennes homog\'{e}nes''},
Ann. Global Anal. Geom. {\bf 27}, 3 (2005) 201 (Eng. ver.: \emph{``Homogeneous nearly K\"ahler manifolds''}, arXiv:math/0612655)
\bibitem{Witten}
  E.~Witten,
  \emph{``Topological Sigma Models''},
  Commun.\ Math.\ Phys.\  {\bf 118} (1988) 411.
\bibitem{ZaremboSym}
  K.~Zarembo,
  \emph{``Strings on Semisymmetric Superspaces''},
  JHEP {\bf 1005} (2010) 002,
  arXiv:1003.0465
\bibitem{Kounnas}
  D.~Israel, C.~Kounnas, D.~Orlando and P.~M.~Petropoulos,
  \emph{``Heterotic strings on homogeneous spaces''},
  Fortsch.\ Phys.\  {\bf 53} (2005) 1030,
  hep-th/0412220
\end{thebibliography}
\end{document}